\documentstyle[12pt,aaspp4]{article}

\lefthead{Yasuhiro Shioya \& Kenji Bekki}
\righthead{Spectroscopic evolution of galaxies}

\begin{document}
\title{Spectroscopic evolution of dusty starburst galaxies
in  distant clusters}

\author{Yasuhiro Shioya} 
\affil{Astronomical Institute, Tohoku University, Sendai, 980-8578, Japan} 

\and

\author{Kenji Bekki and Warrick J. Couch}
\affil{School of Physics, University of New South Wales, 
Sydney 2052, Australia}

\begin{abstract}
By using a one-zone chemical and spectrophotometric evolution
model of a disk galaxy undergoing a dusty starburst,
we investigate, numerically, the optical spectroscopic properties  
in order to explore galaxy evolution in distant clusters.
We adopt an assumption that the degree of dust extinction (represented by
$A_V$) depends on the ages of starburst populations in such a way that
younger stars have larger $A_V$ (originally referred to as selective 
dust extinction by Poggianti \& Wu 2000). In particular, we investigate
how the time evolution of the equivalent widths of [OII]$\lambda$3727 and 
H$\delta$ is controlled by the adopted age dependence. This leads to
three main results: 
(1)\,If a young stellar population (with the age of $\sim$ $10^6$ yr)
is more heavily obscured by dust than an old one ($>$ $10^8$ yr),
the galaxy can show an ``e(a)'' spectrum characterized by strong H$\delta$
absorption and relatively modest [OII] emission. 
(2)\,A dusty starburst galaxy with an e(a) spectrum can evolve into a
poststarburst galaxy with an a+k (or k+a) spectrum 0.2\,Gyr after the
starburst and then into a passive one with a k-type spectrum 1\,Gyr after
the starburst. This result clearly demonstrates an evolutionary
link between galaxies with different spectral classes (i.e.,
e(b), e(a), a+k, k+a, and k).
(3)\,A dusty starburst galaxy can show an a+k or k+a spectrum even
in the dusty starburst phase if the age-dependence of dust extinction is 
rather weak, i.e., if young starburst populations with different ages
($\le$ $10^7$ yr) are uniformly obscured by dust. 

\end{abstract}


\section{Introduction}

Spectroscopic properties of galaxies have provided valuable information 
on different aspects of galaxy evolution, such as star formation
histories along the Hubble sequence, chemical evolution of gas and stars,
and the physical conditions associated with AGN and starburst activity in
the central regions of galaxies (e.g., Kennicutt 1998, 1999). 
In particular, spectroscopic studies of galaxies at optical wavelengths
have played vital roles not only in determining the current and the
past star formation rate but also in deducing the physical mechanisms
relevant to the time evolution of star formation (see Kennicutt 1998
for a recent review). The change in the galactic global star formation
rate with redshift has recently been investigated for a variety of
different distant star--forming objects by many authors (e.g., Madau
1996; Steidel et al. 1999; Ellis 1997). These recent works have suggested
that optical data alone cannot provide a comprehensive view of the star
formation properties of distant galaxies owing to heavy dust extinction.
A major and important challenge is to properly assess the extent to which
the stellar light is affected by dust extinction. Future near-infrared
spectroscopic studies of the redshifted H$\alpha$ emission from distant, 
dust-obscured galaxies is crucial to revealing the degree of extinction
and reddening and thereby clarifying the real star formation histories of
these galaxies (Kennicutt 1999). 

Detailed spectroscopic studies of distant rich clusters (Dressler \& Gunn
1983, 1992; Couch \& Sharples 1987; Poggianti et al. 1999) have provided 
important clues as to the nature and evolution of galaxies in these
dense environments. In particular, Dressler \& Gunn (1983, 1992) discovered 
galaxies with no detectable emission lines but with very strong Balmer
line absorption (the so-called ``E+A'' galaxies) and concluded
that such galaxies were ``poststarburst'' objects seen within $\sim
1$\,Gyr of a severely curtailed burst of star formation. Furthermore,
Couch \& Sharples (1987) made the first attempts to establish an
evolutionary link between star-forming, poststarburst, and passively
evolving galaxies observed in distant clusters. 

These attempts have been progressed significantly further by Dressler et
al. (1999) and Poggianti et al. (1999) through their detailed modeling of
an extensive data base of spectroscopic observations they obtained in 10
rich clusters in the range $0.37\leq z\leq 0.56$. The wealth of data
allowed them to delineate the different spectral classes pertinent to
distant cluster populations much more clearly, with three distinct classes
of emission line object [e(a), e(b), e(c)] being identified on the basis
of the strength of [OII]$\lambda$3727 emission and H$\delta$ absorption, 
two poststarburst classes (a+k and k+a) identified amongst galaxies
with no detectable [OII] emission but enhanced H$\delta$ absorption, and
a passive ``k'' class where neither of these features had any prominence. 
Attempts were made to identify the evolutionary pathways between these
different classes of object in order to identify the most likely 
progenitor to the sizeable population of poststarburst galaxies. 
Galaxies with e(a) spectra characterized by strong Balmer absorption
[EW(H$\delta)>4$\AA] and relatively modest [OII] emission, were previously
interpreted (in the context of dust-free models) as poststarburst galaxies
with some residual star formation (Couch et al. 1998). However, Poggianti et 
al's detailed spectral analysis and comparison with Liu \& Kennicutt's
(1995) sample of spectra for present-day merging galaxies suggested that 
the e(a) types were more likely dusty starburst galaxies. This suggestion
has subsequently been reinforced by the observation that a significant
number of e(a) galaxies are either luminous infrared galaxies (LIRGs) or 
ultra-luminous infrared galaxies (ULIRGs), objects which are often dominated 
by  dusty starbursts (Poggianti \& Wu 2000).  

Independent evidence for the presence and importance of dust
in determining the observed properties of distant cluster galaxies
has been presented by other authors. Smail et al. (1999) found that 5 out
of 10 galaxies classified as either a+k or k+a types by Dressler et
al. (1999) in Abell~851 (at $z=0.42$) were detectable in deep 1.4\,GHz
continuum maps obtained with the VLA. With this radio emission most likely
attributable to starburst activity, the lack of any such signature in the
optical spectrum could only imply that it must be totally obscured by
dust. Further to this, Owen et al. (1999) investigated the radio
properties of galaxies in a  rich cluster at z $\sim 0.25$ (A2125)
which had large blue galaxy fractions (0.19)
and found that 
the optical line luminosities (e.g., H$\alpha$+[NII]) 
were often weaker than one would expect for the star formation rates
implied by the radio emission. They accordingly suggested that
dust obscuration, which is larger than is usually found locally, hides most 
of optical light from the star-forming regions.
 
Although these observational results have identified the possible
importance of dust and dusty starbursts in distant cluster populations,  
little theoretical work has been done to address the formation and evolution of 
such galaxies. The purpose of this paper is to report the results of a 
{\it numerical} investigation of the spectroscopic evolution of dusty
starburst galaxies and possible evolutionary links between galaxies of
different spectral class. Using a one-zone chemical and spectrophotometric 
evolution model, we consider in particular, the degree to which dust
extinction (represented by $A_V$) depends on the age of the starburst
populations and thereby try to clarify the time evolution in the 
equivalent width of the H$\delta$ absorption -- and [OII] emission -- line
features. The main questions that we address are: 
(1)\,When and how does a disk galaxy hosting a dusty starburst develop
an e(a)-type spectrum? (2)\,What are the requisite physical conditions
for the formation of e(a) galaxies? (3)\,Can a dusty galaxy show an 
a+k or k+a spectrum even in its starburst phase? (4)\,Are there any
evolutionary links between the e(a), e(b), e(c), a+k, k+a, and k spectral
types?

A key feature of our approach is to adopt the {\it selective
dust extinction} framework originally proposed by Poggianti \& Wu (2000)
whereby the extinction suffered by a stellar population is inversely
proportional to its age with the youngest populations having the 
greatest extinction. We confirm that such age-dependent extinction 
is essential to reproducing the spectral properties of the putative
dusty starburst population in distant clusters.

The layout of this paper is as follows: In \S 2, we summarize the
numerical models used in the present study and describe briefly the
methods for deriving spectral energy distributions (SEDs) affected by
internal dust extinction. In \S 3, we present our numerical results on
the time evolution of dusty starburst galaxies in the 
EW([OII])-EW(H$\delta$) plane. In \S 4, we discuss the possible 
evolutionary connections between the different spectral types 
(e.g., e(a), e(b), and k+a) in the context of our dusty starburst models. 
The conclusions of the present study are  given in \S 5.
In the present paper,
each spectral class is defined as follows: e(a) for galaxies
with 5 \AA $<$ EW([OII]) $\le$ 40 \AA and 4 \AA $<$ EW(H$\delta$),
e(b) for  40  \AA $<$ EW([OII])  and  EW(H$\delta$) $\le$ 4 \AA, 
e(c) for  5 \AA $<$ EW([OII]) $\le$ 40 \AA and  EW(H$\delta$) $\le$ 4 \AA,
a+k  for  EW([OII]) $\le$ 5 \AA and  7 \AA $<$ EW(H$\delta$),
k+a  for  EW([OII]) $\le$ 5 \AA and  3 \AA $<$ EW(H$\delta$)  $\le$ 7 \AA,
k  for  EW([OII]) $\le$ 5 \AA and  EW(H$\delta$)  $\le$ 3 \AA.
 
\section{Model}

We adopt a one-zone chemical and spectrophotometric evolution model 
of a disk galaxy with a starburst, and thereby investigate 
the time evolution of the line strength of  H$\delta$ absorption and
that of [OII] emission, both of which are important properties
for classifying the spectral types of galaxies in distant clusters
(Dressler et al. 1999). The present model is essentially the same as that
adopted in our previous studies (Shioya \& Bekki 1998, 2000), with the
details of our treatment of chemical evolution  given in Shioya \& Bekki
(1998). Accordingly, we describe only briefly the present spectrophotometric 
model. 

One of the main  differences between our model and previous
ones (e.g., Couch \& Sharples 1987; Newberry et al. 1990; 
Charlot \& Silk 1994; Jablonka \& Alloin 1995; 
Poggianti \& Barbaro 1996; Poggianti et al. 1999) is that we explicitly
include the effects of dust on both the photometric and spectroscopic
properties of stellar populations as also the gaseous emissions.
With this new development to our code, we can clearly demonstrate how
the dust extinction affects the time evolution of the line strengths
of  H$\delta$ and [OII] (EW(H$\delta$) and EW([OII])) in dusty starburst
galaxies.

We follow the chemical evolution of galaxies by using the model 
described in Matteucci \& Tornamb\`{e} (1987) which includes 
metal-enrichment processes of Type Ia and II supernovae (SNIa and SNII). 
We adopt the Salpeter initial mass function (IMF), $\phi(m) \propto
m^{-1.35}$, with upper mass limit $M_{\rm up}=120M_{\odot}$ and lower
mass limit $M_{\rm low}=0.1M_{\odot}$. 
We calculate the photometric properties of galaxies as follows: 
The monochromatic flux of a galaxy with age $T$, $F_{\lambda}(T)$, 
is described as:  
\begin{equation}
F_{\lambda} (T) = \int_0^T F_{\rm SSP,\lambda}(Z,T-t) \psi(t) dt, 
\end{equation}
where $F_{\rm SSP,\lambda}(Z,T-t)$ is the monochromatic flux of 
a single stellar population with age $T-t$ and metallicity $Z$, and 
$\psi(t)$ is the time-dependent star formation rate described below. 
In the present study, we use the spectral library GISSEL96 which is the 
most recent version of that published by Bruzual \& Charlot (1993). 

The star formation history of a disk galaxy is characterized by three epochs.  
The first is the epoch of galaxy formation ($T_{\rm form}$) during which
the interstellar gas is consumed by star formation at a moderate rate. 
In the present study, $T_{\rm form}$ is 0 for all models. The second is
$T_{\rm sb}$ at which the starburst begins in the disk. Here $T_{\rm sb}$
is considered to be a free parameter. The third is $T_{\rm end}$ at which
$all$ star formation ceases 
is defined as the epoch at which the stellar mass fraction 
reaches a value $f_{\rm final}$ in our models; $f_{\rm final}$ is also
assumed to be a free parameter which controls the duration of
starburst. Throughout the evolution of disk galaxies, the star formation
rate is assumed to be proportional to the gas mass fraction ($f_g$):  
\begin{equation}
\psi(t)=kf_g,
\end{equation}
where $k$ is a parameter which controls the star formation rate. 
This parameter $k$ is given as follows: 
\begin{equation}
k = \left\{
\begin{array}{lcl}
k_{\rm disk} \; \; \; & & {\rm for} \; \; 0 \le T < T_{\rm sb}, \\
k_{\rm sb}   & & {\rm for} \; \; T_{\rm sb} \le T < T_{\rm end}, \\
0            & & {\rm for} \; \; T_{\rm end} \le T.
\end{array}
\right.
\end{equation}
In the following, we refer to these three different epochs as the  
``prestarburst'',  ``starburst'', and ``poststarburst'' phases, 
respectively.  

The value of the $k_{\rm disk}$ parameter, which controls the star
formation histories of disk galaxies, is suggested to vary for different
Hubble types (e.g., Arimoto, Yoshii, \& Takahara 1992). For example,
$k_{\rm disk}$ = 0.225 ${\rm Gyr}^{-1}$ corresponds to a plausible star 
formation rate for Sb disks (e.g., Arimoto et al. 1992). In comparison, 
observations of starburst galaxies (e.g., Planesas et al. 1997) would
suggest that $k_{\rm sb}$ is an order of magnitude larger than 
$k_{\rm disk}$ (i.e., $k_{\rm sb}$ = $2.25 \; {\rm Gyr}^{-1}$). 
By changing the value of $k_{\rm sb}$, we can investigate how the strength
of a starburst influences the evolution in spectral type within a
dusty galaxy.

We adopt the selective dust extinction model investigated in detail
by Shioya \& Bekki (2000). In this model, the amount of dust
extinction associated with the starburst is assumed to depend on the age
of stellar population: 
\begin{equation}
A_V=A_{V,0} \exp \{ (T-t)/\tau \} \; \; \; 
{\rm for} \; \; T_{\rm sb} < T < T_{\rm end}, 
\end{equation}
where $A_{V,0}$ and $\tau$ are free parameters which control the
the extinction at any given time during starburst.
Based on the monochromatic flux derived in 
equation (1) and the value of $A_V$ given by equation (4),   
we calculated a `dust-absorbed' SED for galaxies of different age $T$
using the extinction law inferred  by Calzetti et al. (1995) and adopting
the so-called screen model. To derive fluxes for various gaseous emission 
lines (H$\delta$ and [OII]), we first calculate the number of Lyman
continuum photons, $N_{\rm Ly}$, by using a non-absorbed SED.
If all the Lyman continuum photons are consumed in ionizing the
surrounding gas, the luminosity of H$\beta$ can be calculated according to
the relation: 
\begin{equation}
L({\rm H}\beta) ({\rm erg \; s^{-1}}) = 4.76 \times 10^{-13} N_{\rm Ly} ({\rm s^{-1}})
\end{equation}
(Leitherer \& Heckman 1995). 
To calculate luminosities of other emission lines, e.g., [OII] and
H$\delta$, we use the luminosity ratios relative to H$\beta$ tabulated 
in PEGASE (Fioc \& Rocca-Volmerange 1997), which are calculated 
for an electron temperature of 10,000\,K and an electron density of 
1\,cm$^{-3}$. The strength of absorption lines are derived using 
GISSEL96 (Bruzual \& Charlot 1993). Thus the SEDs derived here comprise
the following components: stellar continuum, gaseous emission, and stellar
absorption. 

\subsection{Main points of analysis}

Shioya \& Bekki (2000) have already demonstrated the importance of
selective dust extinction in reproducing e(a) type galaxy spectra by
comparing the model results with and without dust extinction.
Furthermore, they pointed out that a dusty starburst galaxy
with selective dust extinction evolves from an e(b), to an e(a), 
to an a+k, to a k+a, and finally to a k type galaxy. Since they have
already indicated the fundamental role of the selective dust extinction
in spectroscopic evolution of dusty starburst galaxies, we detail here the
parameter dependence of the spectroscopic evolution in the 
EW([OII])-EW(H$\delta$) plane. To demonstrate the importance of each
of the six free parameters, $A_{V,0}$, $\tau$, $T_{\rm sb}$, $k_{\rm
disk}$, $k_{\rm sb}$, and  $f_{\rm final}$, we firstly compute a
fiducial model (Model 1) with $A_{V,0}$ = 5.0 mag, $\tau$ = $1.0
\times 10^6$ yr, $T_{\rm sb}$ = 7.64 Gyr, $k_{\rm disk}$ = 0.225
Gyr$^{-1}$, $k_{\rm sb}$ = 2.25 Gyr$^{-1}$, and $f_{\rm final}$ = 0.95.
Model 1 can then be compared with a model without dust extinction (Model
2), to show the typical behavior of selective dust extinction in the
evolution of starburst galaxies on the EW([OII])-EW(H$\delta$) plane.  
The dependence of the evolution on the six parameters, $A_{V,0}$, $\tau$,
$T_{\rm sb}$, $k_{\rm disk}$, $k_{\rm sb}$, and  $f_{\rm final}$ is then
investigated via the computation of 17 models with a range of parameter
values:
\begin{description}
\item Model 1, 3, 4,and 5 with different  $A_{V,0}$,
\item Model 1, 6, and 7 with different $\tau$,
\item Model 1, 8, 9, 10, and 11 with different  $T_{\rm sb}$,
\item Model 1, 12, and 13 with different $k_{\rm disk}$, 
\item Model 1, 14, and 15 with different $k_{\rm sb}$, 
\item and Model 1, 16, and 17 with different $k_{\rm final}$. 
\end{description}
The parameter values are given in Table 1. We assume that the initial gas
mass of the model galaxy is $10^{11}M_{\odot}$.

\placefigure{fig-1}
\placefigure{fig-2}
\placefigure{fig-3}
\placefigure{fig-4}

\section{Results}

\subsection{The fiducial model}

Figure 1 shows the star formation history and the color evolution (in $B-V$) 
for the fiducial model (Model 1) with $k_{\rm disk}$ = 0.225 ${\rm Gyr}^{-1}$
(corresponding to Sb--type star formation history). During the starburst,
the $B-V$ color becomes rather blue  (0.2 mag at $T$ =  8.56 Gyr) due 
to the formation of massive stars. However, the color is not so blue
compared with the model without selective dust extinction (Model 2) 
because of the heavy dust obscuration suffered by the young stellar
population in the fiducial model. This result implies that selective dust
extinction affects the global colors as well as the spectral characteristics
of starburst galaxies.

Figure 2 shows the evolution of the  emission and absorption lines 
by plotting their strength at $T$ = 1, 7.64, 8.56, 8.58, and 13.0 Gyr for
the fiducial model. Before the starburst ($T$ $<$ 7.64 Gyr), the model
clearly shows a number of strong emission lines, whereas the absorption
lines (e.g., the H$\delta$ absorption line) are rather weak as a result of 
the continuous and moderate star formation. During the starburst  (7.64
$\le$ $T$ $\le$ 8.58 Gyr), the emission lines do not become so remarkably
strong; nevertheless, the star formation rate becomes more than a factor
of five larger than that of the prestarburst phase. This is because 
the ionizing photons from very young OB stars are very heavily obscured by
the dust. After the starburst ($T$ $>$ 8.58 Gyr), the emission lines
become even less remarkable due to the death of the young, massive stars
and the intermediate--aged ones as well. On the other hand, the stellar
absorption lines (e.g., H$\delta$) become prominent after the starburst
($T$ $>$ 8.58 Gyr) owing to the presence of an older stellar population. 

Figure 3 describes the time evolution of EW([OII]) and EW(H$\delta$), which 
clearly demonstrates the importance of the selective dust extinction in
the spectroscopic evolution of dusty starburst galaxies. The equivalent
width of [OII] does {\it not} increase and is well below 40 \AA~ 
during the starburst. The equivalent width of H$\delta$ becomes rapidly
smaller just after the starburst begins, although the H$\delta$ line can
still be observed as an absorption line. A few $\times 10^8$ years after 
the starburst begins, the equivalent width of H$\delta$ becomes larger
than 4\,\AA. Consequently, this model reproduces the properties of the
e(a) class with EW(H$\delta$) $>4$ \AA and relatively modest [OII] emission, 
EW([OII])$>5$  \AA; the location in the EW([OII])-EW(H$\delta$) plane
is shown in Figure 4. The physical reason for the successful reproduction
of the model is clearly described in our previous paper (Shioya \& Bekki
2000). The time scale during which the model is located within the
e(a) region is estimated to be $\sim 0.7$\,Gyr. These results thus
confirm the earlier suggestion of Poggianti \& Wu (2000) of the vital
role of selective dust extinction in reproducing e(a) spectra.

\placefigure{fig-5}
\placefigure{fig-6}
\placefigure{fig-7}

\subsection{Parameter dependences}

Figure 5 shows the time evolution of spectroscopic properties in the
EW([OII])-EW(H$\delta$) plane for models with different values of
$A_{V,0}$ (1 $\le$ $A_{V,0}$ $\le$ 10.0 mag for Model  2, 3, 4, and 5). 
Since parameters other than $A_{V,0}$ are fixed at the same values
adopted for the fiducial model, this figure demonstrates how the evolution
of dusty starburst galaxies in the EW([OII])-EW(H$\delta$) plane differs
according to the amount of dust extinction. As can be seen in the figure,
the locus on the plane is appreciably different between the models,
which implies that the degree of dust extinction (represented by
$A_{V,0}$) is an important influence on the observed spectroscopic
evolution. 
During the starburst, the model with the larger $A_{V,0}$ shows weaker
[OII] emission, weaker H$\delta$ emission, and stronger H$\delta$
absorption. Thus the model with the largest $A_{V,0}$ (e.g., Model 5 with
$A_{V,0}$ = 10 mag) is located towards the bottom--right of the 
e(a) region during the dusty starburst.

Figure 6 describes the evolution in the EW([OII])-EW(H$\delta$) plane 
for the models with different values of $\tau$ ($1\times 10^6\le\tau\le 1
\times 10^7$\,yr for Model  1, 6, and 7). Parameters other than $\tau$
are fixed at the same values of the fiducial model. This figure clearly
demonstrates that the time-scale during which young massive stars
are heavily obscured by dust also governs the spectroscopic
evolution. For the parameter range adopted here, a larger number of
young massive stars with different ages are obscured by dust for the model
with larger $\tau$, i.e., the age-dependent dust extinction is weaker and
thus young stars with different ages ($<10^7$\,yr) are more uniformly
obscured by dust. Therefore, the model with larger $\tau$ shows lower
EW([OII]) and higher EW(H$\delta$) and, as a result, is located in the
bottom-right hand region of the EW([OII])-EW(H$\delta$) plane during
the starburst. We stress here that for the model with larger $\tau$  (= $3
\times 10^6$\,yr and $1\times 10^7$\,yr; Model 6 and 7, respectively),
the time-scale during which the galaxies show an e(a)-type spectrum is
appreciably shorter compared to the fiducial model. Furthermore, the
galaxies show a k+a/a+k spectral signature in the late phase of the 
starburst. This result suggests that not only the galaxies with e(a) spectra
but also those with k+a/a+k spectra are possible dusty starburst candidates 
in distant clusters. 

For the models with $\tau >1 \times 10^7$\,yr, not only young massive
stars (e.g., OB stars) -- which emit most of the ionizing photons -- but
also late B and A stars are strongly affected by dust extinction. Hence
a galaxy with a $\tau$ value of this size does not show strong H$\delta$
absorption during the starburst and poststarburst phases. Therefore the
galaxy can be located in k+a region even in the dusty starburst phase.
This can be contrasted with models with $\tau <1 \times 10^6$\,yr, where
the emission lines from ionized gas regions around massive stars are not
obscured so heavily by dust. As a result of this, even if we adopt a
rather large value for $A_{V,0}$ (i.e., more than 10 mag), neither the
[OII] nor H$\delta$ lines are weakened. In this case, a dusty starburst
galaxy shows an e(b) spectrum during its starburst phase.
 
These results on the $\tau$--dependences within our models confirms that
age-dependent dust extinction is a crucially important factor for
determining the spectroscopic properties of dusty starburst galaxies.
The present one-zone model, however, cannot determine with any great
certainty what controls the age-dependence (i.e., the parameter value
of $\tau$) owing to the limitations of the model. We further discuss this
problem in a later section.

Figure 7 describes how the star formation history and the evolution of
spectroscopic properties on the EW([OII])-EW(H$\delta$) plane are
controlled by the parameter $T_{\rm sb}$ in dusty starburst galaxies.
For models with a smaller $T_{\rm sb}$, the starburst occurs at earlier
epochs and the magnitude of the starburst is larger.
This  is because there is more
gas to fuel starbursts  at earlier times.
The evolution of galaxies on the EW([OII])-EW(H$\delta$) plane is not so
remarkably different between the models with different $T_{\rm sb}$
(=4.46, 6.93, 7.64, 8.47, and 9.45 Gyr: corresponding to Model 8, 9, 1,
10, and 11, respectively). The model with smaller $T_{\rm sb}$ (e.g.,
Model 9), which represents a higher redshift starburst galaxy, shows
discernibly lower EW([OII]) and higher EW(H$\delta$) and thus is located
at the bottom-right of the e(a) region during the starburst. This is
primarily because a larger number of young massive stars are created and
then obscured preferentially in the model with smaller $T_{\rm sb}$.
These results firstly imply that irrespective of the epoch at which a
dusty starburst occurs, disk galaxies can exhibit an e(a) spectrum during
this phase if the extinction parameters are within certain reasonable
ranges. Secondly, they imply that the higher redshift dusty starburst
galaxies are more likely to be located to the right and to the bottom
of the e(a) region in the EW([OII])-EW(H$\delta$) plane.  It would be
very interesting if future spectroscopic observations could confirm
this predicted redshift-dependent behavior in dusty starburst galaxies.

Figure 8 describes how the evolution of starburst galaxies in the
EW([OII])-EW(H$\delta$) plane depends on their prior star formation
histories. The three models with Sa (Model 13), Sb (Model 1), and Sc 
(Model 12) type star formation histories all pass through the e(a) region
during the starburst, which implies that a dusty starburst galaxy can show 
an e(a) spectrum irrespective of its longer-term activity in its quiescent
state.

Figure 9 shows the dependence of the spectroscopic evolution on the
parameter $k_{\rm sb}$ controlling the strengths of starbursts
in Sb type disk galaxies. Since parameters other than $k_{\rm sb}$ 
are the same between these models (with  $k_{\rm sb}$ = 1.0, 2.25,
and 5.0 Gyr$^{-1}$, corresponding to Models 14, 1, and 15, respectively),
this figure describes how the strength of a starburst in a galaxy
determines the location of the dusty galaxy in the EW([OII])-EW(H$\delta$) 
plane. For the adopted parameter range ($1\le k_{\rm sb}\le 5.0$\,Gyr$^{-1}$),
the models show e(a) spectra during starbursts, although the location of
each model in the e(a) region is different between these models: The
models with larger $k_{\rm sb}$ (i.e., those with a stronger starburst)
show stronger EW(H$\delta$) and are thus located to the right of the 
e(a) region. The two models with $k_{\rm sb} = 2.25$ and 5\,Gyr$^{-1}$
evolve from e(a) to a+k galaxies whereas the model with $k_{\rm sb} = 1.0$, 
corresponding to a galaxy with a rather weak and longer starburst, evolve
from e(a) to k+a without passing through the a+k region. This result
implies that dusty starburst galaxies that have undergone a rather weak 
starburst, are more likely to exhibit a k+a spectrum straight after the
starburst. Clearly the {\it strength} of the starburst is important to
the evolution of a dusty starburst in the EW([OII])-EW(H$\delta$) plane. 

Figure 10 demonstrates how the evolution is dependent on the duration of
the starburst. There is about 1.8\,Gyr difference in the starburst
duration between the models, with $f_{\rm final}$ = 0.90 (Model 16), 0.95
(Model 1, fiducial model), and 0.99 (Model 17). All three of the models
pass through the e(a) region during the starburst, albeit at different
locations. The models with smaller $f_{\rm final}$ values (Model 1 and
14) -- corresponding to those with the shorter starburst durations
-- evolve from e(a) into a+k and then into k+a types during the starburst
and poststarburst phases. In contrast, the model with the largest
$f_{\rm final}$ value and thus the longest starburst duration, evolves 
directly from an e(a) to a k+a type without passing through the a+k
spectral phase. (This dependence on $f_{\rm final}$ is strikingly
similar to that on $k_{\rm sb}$.) This result implies that a dusty
starburst galaxy with the longer starburst duration is more likely to show
the k+a spectra in the poststarburst phase.

To summarize the above, 
weak [O II] equivalent width during the
starburst is favoured by large dust obscuration, obscuration which is slow to
clear relative to the characteristic age of the starburst.
Furthermore, the location and movement within the
EW([OII])-EW(H$\delta$) plane of a dusty starburst galaxy during both the
starburst and poststarburst phases has been demonstrated to be dependent
on a number of parameters: $A_{V,0}$, $\tau$, $T_{\rm sb}$, $k_{\rm sb}$,
$k_{\rm disk}$, and  $f_{\rm final}$. In turn, these parameters could
depend on the relative spatial distribution of gas and stars, the radial
variation in the metallicity and density of the gas, the dynamical
evolution of gas and stars, and the effectiveness of supernovae feedback
in the evolution of the interstellar medium for a dusty starburst galaxy.
The one-zone model adopted here is, by its very nature, unable to 
determine the extent to which these important factors (e.g., the relative
spatial distribution of gas and stars) are closely associated with the origin 
of the six parameters.  This means that although the present study
can indicate which of those parameters are important to the spectroscopic 
evolution of dusty starburst galaxies, it cannot so clearly reveal the
origin of the spectroscopic properties of the galaxies. Thus future
studies, in which the above five parameters are not treated as free
parameters but as ones that can be predicted from a certain theoretical
calculation are required to understand more clearly the origin of each
spectral type (e.g., e(a)) in the context of distant cluster galaxy
evolution.

\placefigure{fig-8}
\placefigure{fig-9}
\placefigure{fig-10}


\section{Discussion}

\subsection{Origin of selective dust extinction}

Calzetti et al. (1994) and Mayya \& Prabhu (1996).
have already proposed the idea that the older populations can be located at the outer part of
molecular clouds and thus do not suffer so severely from dust extinction.
Poggianti \& Wu (2000) furthermore pointed out that stellar populations with
the ages between a few $10^7$ and 1.5 $\times 10^9$\,yrs, which are
responsible for the strong Balmer line absorption, have had sufficient 
time to drift away from the dusty molecular clouds where they were born.  
This selective obscuration of young stellar populations probably 
works effectively wherever a starburst occurs within molecular clouds.
Both Shioya \& Bekki (2000) and the present study have demonstrated that
if younger starburst populations are more heavily obscured by dusty
gas than old ones in a galaxy, the galaxy shows an e(a)-type spectrum. 
This was first suggested by Poggianti \& Wu (2000), who also proposed
that if the location and thickness of dust depended on the age of
the embedded stellar populations in a starburst galaxy, then the effects
it would have on its spectroscopic properties would be quite remarkable.

Why is it then that only some fraction of dusty starburst galaxies show
e(a) spectra? Moreover, when observing them in distant clusters, it should
not be forgotten that their spectrum is an integrated one, comprising a
luminosity-weighted `mean' of the light emitted over a large region of the
galaxy. Thus while the selective obscuration suggested by Poggianti \& Wu
(2000) is clearly important, it is the {\it global} (100--1000\,pc) rather 
than the local ($\sim 10$\,pc) processes that need to be investigated in
in detail to explain the origin of the selective extinction.

We point out here that a difference in the degree of dust extinction
between the central ($<$ a few 100 pc) and outer ($\geq 1$\,kpc) regions
of a galaxy ($\sim$ 1 kpc) could produce the above age-dependent
extinction effect. To be more specific, if a young stellar population
is formed in a secondary starburst in the central region where there 
are greater quantities of dust and hence more dust extinction, its
contrast in properties to the older and less obscured populations in the
outer regions of the galaxy will make the extinction appear as though it
is selective. This idea that the {\it age} of the stellar population
varies globally with radius is not unreasonable and unrealistic, given
that secondary starbursts occur preferentially in the central parts of
galaxies due to the efficient inward transfer of gas driven by
non-axisymmetric structure (such as stellar bars) and galaxy interaction
and merging (Barnes \& Hernquist 1992;  Mihos 
\& Hernquist 1995). Furthermore, 
numerical simulations of the gaseous and stellar distribution in merging 
disk galaxies (which trigger a dusty starburst) have demonstrated that 
the central young stellar component formed by a nuclear starburst, is more 
heavily obscured by dust than the outer, older component initially 
located in the merger progenitor disks (Bekki, Shioya, \& Tanaka 1999). 
Bekki \& Shioya (2000) also demonstrated that since younger stellar
populations are more centrally concentrated and the gas density is
higher in the inner region, the mean gas density around stars (which
is critically important for $A_V$) depends strongly on stellar age in
such a way that the density around younger populations is higher than
that around older populations. Therefore we strongly suggest that
galaxies which undergo central dusty starbursts are more likely to show
selective dust extinction. 

If the suggested radial dependence of stellar age and dust extinction is
plausible and realistic, are there physical processes operating in 
distant rich clusters which have an equally realistic chance of causing
it? Physical processes which are possibilities include: tidal galaxy
interaction (Icke 1985;  Noguchi 1987), tidal effects of cluster cores
(Byrd \& Valtonen 1990), galaxy `harassment' (Moore et al. 1996), minor
merging (Mihos \& Hernquist 1994), unequal-mass merging (Barnes
1996; Bekki 1998), and major merging (Barnes \& Hernquist 1991).
Given the high probability of high speed galaxy encounters in the
rich cluster environment, galaxy harassment seems to be the most
promising candidate. However it has the following three disadvantages in
explaining the nature of e(a) galaxies. Firstly, the blue cluster
galaxies of modest luminosity, i.e. in the luminosity regime where galaxy
harassment is effective, show predominantly e(b)-- rather than e(a)--type
spectra (Poggianti et al. 1999). Secondly, there is no clear tendency that
e(a) galaxies are preferentially located within the cores of clusters
(Dressler et al. 1999), where galaxy harassment is most likely to trigger
a starburst. Thirdly, combined spectroscopic and morphological studies of
the galaxies in distant clusters (e.g. Couch et al. 1998) have not revealed
any significant population of starburst objects which has the grossly
deformed spiral arm pattern predicted by Moore et al. (1996). Of course
these observations do not necessarily rule out galaxy harassment as a
mechanism for forming e(a) galaxies; rather, if galaxy harassment is
important, then it will produce only low luminosity galaxies located
near the cluster core. The second observation above would also appear to
rule out tidal effects within cluster cores as being important to
e(a) formation as well. 

A rather high fraction ($\sim$ 50 \%) of low--$z$LIRGs and ULIRGs  --
considered to be ongoing major mergers -- are observed to have e(a) 
spectra (Liu \& Kennicutt 1995, Poggianti \& Wu 2000). The rate of
galaxy merging is found to increase very steeply with redshift 
($\propto {(1+z)}^{6 \pm 2}$; van Dokkum et al. 1999) and, furthermore, 
there is some evidence that this trend is seen amongst e(a) galaxies 
(Dressler et al. 1999; Poggianti et al. 1999). On the other hand, 
the e(a) galaxies found in the distant cluster samples are predominantly
{\it disk} galaxies, which makes it hard to reconcile them with being
the products of major mergers, which tend to destroy any disk structure. 

We suggest, therefore, that minor and unequal-mass galaxy merging, which 
is different from major merging in that it can leave disk systems 
intact, is a more likely mechanism for reproducing the fundamental
properties of e(a) galaxies. Mihos \& Hernquist (1994) demonstrated that
minor merging can excite non-axisymmetric structure in a late-type disk,
trigger nuclear starbursts, and transform the disk into an early-type
system such as an Sa or S0 galaxy. Furthermore, unequal-mass merging
between two disks has been demonstrated to create a disk with structure
and kinematics strikingly similar to that of an S0 (Barnes 1996). On
that note, Bekki (1998) also found that such merging can trigger nuclear
starburst, form a very gas-poor disk, and create bulge with characteristics 
remarkably similar to that of typical S0s. Hence as well as contributing 
to the formation of dusty nuclear starbursts and thus to the radial 
dependence of dust extinction, this more minor type of merging may also 
provide an evolutionary link between late-type disks, e(a) galaxies, and
early-type disks (S0s) and thus account for the observed decrease in the
cluster S0 fraction with redshift (e.g., Dressler et al. 1996). 
An early-type disk galaxy formed by a minor/unequal-mass merger, will 
experience a poststarburst phase (with k+a/a+k spectra), and should 
contain a bulge remarkable for its young stellar population and thus
bluer colors (Bekki 1998). Accordingly, a key observational test for
assessing the relative importance of minor/unequal-mass merging in the
formation of e(a) galaxies with selective dust extinction, is to
investigate the number fraction of disk galaxies with k+a/a+k spectra and
prominent $blue$ bulges in distant clusters.

\subsection{On the origin of less luminous starburst galaxies with
e(b) spectra}

We now focus on the origin of the difference between the e(a) and
e(b) spectral classes: Why do some starburst galaxies show e(b) spectra
while others show e(a) spectra? We have demonstrated that the spectral
type of a dusty starburst galaxy has a basic dependence $A_{V,0}$ in that 
it will exhibit an e(a) or e(b) spectrum if $A_{V,0}$ is large or small, 
respectively. 

Poggianti et al. (1999) modeled the time evolution of the global colors
and luminosities of distant cluster galaxies by using a stellar population
synthesis code and found that the e(b) galaxies were likely to evolve 
into sub--luminous systems once the starburst had ceased. They also found 
that the HII region metallicities of the e(b) galaxies are significantly
lower than those of galaxies in other spectral classes, and that
these low metallicities were consistent with them being low-luminosity
galaxies. These observational results suggest that one of the physical
parameters that can govern the difference in spectral class is galactic
luminosity. Given our result that the spectral class of a dusty starburst
galaxy depends on $A_{V,0}$, this would imply that the luminosity and  
$A_{V,0}$ are somehow related. How can this be? One possibility is that 
for the more luminous galaxies, the thermal feedback of supernovae
explosions that occur during starbursts is simply not strong enough to
blow off some fraction of the interstellar gas surrounding the young
stars and thus to modify the spatial distribution of the gas. The key
factor here is the deeper gravitational potential well of the galaxy which
prevents the gas from escaping the galaxy. As a result of this,
the mean gaseous density (which is an important factor in determining  
$A_{V,0}$) continues to remain high, as also its dust extinction.
Consequently, the galaxy can show an e(a) spectrum.

For a less luminous galaxy, most of the dusty interstellar gas can be
blown away by the thermal feedback effect of the supernovae, as a result
of the shallower gravitational potential. Consequently, the density of
dusty gas around the starburst populations is reduced as also the degree
of dust extinction. The galaxy thus shows an e(b) spectrum. Hence we
suggest that owing to the difference in the effectiveness of supernovae
feedback in galaxies with different luminosities, galactic luminosity can
determine the degree of dust extinction and optical spectral type.

Another possibility for the origin of e(b) galaxies is that they 
are normal star-forming systems but with a significantly lower
metallicity. Although Poggianti et al. (1999) argued that e(b) galaxies
must be starburst objects because a spiral-like star formation 
history was not able to produce strong enough [OII] emission, viz.
EW([OII])$=40$\AA. However, this was based on models with solar
metallicity and indeed Poggianti et al. acknowledged that the relation 
between SFR and [OII] strength can change with metallicity. In estimating
the strength of the [OII] flux for a galaxy with a given star formation
rate, the H$\beta$ flux is first derived and then used to estimate the
[OII] strength by adopting a certain [OII]/H$\beta$ ratio. Accordingly,
if less luminous and less metal-rich galaxies have a larger [OII]/H$\beta$
ratio, then they will have larger [OII] fluxes for a given star formation
rate. The large [OII] fluxes observed in e(b) galaxies could therefore
be the result of such a metallicity effect rather than due to a high
star formation rate. Indeed, Kobulnicky et al. (1999) have found that
the [OII]/H$\beta$ ratio correlates with oxygen abundance (12+log(O/H))
in such a way that a galaxy with smaller gaseous metallicity
is more likely to show a larger [OII]/H$\beta$ ratio. This observational
result seems to support the above possibility that e(b) are just normal
star-forming galaxies with less metal-rich gas. If this is the case,  
we do not need to consider the reason why there are two different
dusty starburst populations. 

\subsection{The dusty starburst nature of the k+a/a+k population?}

Smail et al. (1999) found that a significant fraction of the k+a
population in Abell~851 ($z=0.42$) appeared as sources in deep
VLA 1.4\,GHz continuum maps of this cluster, and suggested that even 
k+a galaxies -- classified {\it optically} by their lack of [OII] emission
but moderately strong H$\delta$ absorption -- can be systems hosting
and ongoing dusty starburst. Balogh \& Morris (2000) also found examples 
of k+a galaxies with H$\alpha$ emission and, accordingly, concluded that
some fraction of k+a galaxies are likely to be undergoing significant
star formation. These observations have raised the following two
questions: (i)\,Are such k+a galaxies really dusty starburst ones?, and 
(ii)\,What determines the spectral types of dusty starburst galaxies (or,
why are there dusty starburst populations with spectral characteristics,
viz. e(a) and k+a)? Concerning the first question, Smail et al. (1999)
argued that the radio emission from the k+a galaxies most likely had
a starburst origin (being the synchrotron radiation emitted from Type II
supernovae associated with the newly formed population of massive stars)
with star formation rates of the order of 
10--30\,$M_{\odot}$\,${\rm yr}^{-1}$. Their `poststarburst' spectral
signature at optical wavelengths would imply that this star formation
was completely obscured. Although this interpretation is plausible and
consistent with other observational results such as color images,
submillimeter fluxes, and the [OII] emission line strengths for these 
galaxies  (Smail et al. 1999; Poggianti et al. 1999), we point out that
an alternative interpretation is that there are significant spatial
variations in star-forming activity across these galaxies, with the
central regions in a ` poststarburst' phase and the outer regions still
in a `starburst' phase. The galaxy shows a k+a spectrum at optical
wavelengths because the earlier $central$ star formation (which forms the
poststarburst population) occurred more intensively than the $outer$
ongoing star formation: the integrated spectrum is simply dominated by
the light from the  central poststarburst region. At radio wavelengths, 
the flux comes exclusively from the outer star-forming region.
The validity of this interpretation can be assessed observationally, if
the detailed two-dimensional spatial distribution of say H$\alpha$
emission is mapped. Furthermore, we can confirm whether Smail et al.'s
interpretation of these objects is robust by investigating their infrared
fluxes and thereby obtaining more accurate star formation rates.
Thus it is reasonable to say that future H$\alpha$ and infrared observations 
will resolve this first question. 

Concerning the second problem, our present numerical results can provide 
some clues. Our models for starbursts with $\tau$ = $10^6$ and $10^7$
show e(a) and k+a spectra, respectively, which implies that the difference
in $\tau$ yields the spectral type difference. The larger value of $\tau$
means that the age-dependence of dust extinction is weaker and thus 
interstellar dust uniformly obscures the starburst populations of
different ages. Accordingly, dusty starburst galaxies with k+a spectral
types are likely to be formed, if younger stellar populations are not so
preferentially obscured by dust (i.e., the selective dust extinction does
not effectively work). If the difference in $\tau$ is really an important
factor in explaining the origin of the e(a)/k+a dichotomy, the next
question is what physical processes drive the $\tau$ difference.
Answering this question is obviously beyond the scope of this paper,
because $\tau$ is a free parameter of the present one-zone model.
However, we suggest that the difference in spatial distribution
of stars, gas, and dust between dusty starburst galaxies is
responsible for the $\tau$ difference. In our future papers, we will
investigate how the radial distribution of stars and gas affects the
spectral types of dusty starburst galaxies.

\subsection{Future work: Spectrodynamical simulations}

Although the present one-zone model of dusty starburst galaxies
has clarified some essential ingredients of the spectroscopic
evolution of distant cluster galaxies, it has the following three
limitations in analyzing the spectroscopic properties of galaxies.
Firstly, the degree of dust extinction (represented by $A_V$),
which is one of the most important parameters in determining the SEDs of
galaxies, is a free parameter in the one-zone model. $A_V$ is generally
considered to depend on gaseous density, gaseous metallicity, dust
composition, and the relative distribution of gas and stars (e.g.,
Mazzei, Xu, \& De Zotti 1992) and so is not calculated in principle
by the one-zone model. Secondly, the dependence of $A_V$ on the ages of
the starburst populations is assumed in an arbitrary way and fixed for
the one-zone model. The age dependence, which is another important factor
in this context, depends on the relative spatial distribution of dusty
gas and young stars and thus cannot be predicted from the one-zone model. 
Thirdly, the star formation history of a galaxy is freely controlled  
for the one-zone model. The time evolution of the star formation rate is
suggested to be controlled by dynamical processes such as local
gravitational instability in disk galaxies (e.g., Kennicutt 1989, 1998); this
is not addressed by the one-zone model since it is unable to follow the 
dynamical evolution of galaxies. In future work, we suggest that the above 
three properties (e.g., $A_{V}$) in the one-zone model should not be freely  
parameterized but predicted from theoretical or numerical studies so that
spectroscopic properties can be investigated in a more self-consistent
manner.  

The most effective way to remove the above three limitations in
the classical one-zone model of spectrophotometric evolution
is to derive both the detailed distributions of gas, dust, and stars and
the time-evolution of the star formation rate, by performing numerical
simulations of dusty starburst galaxies. We have already developed a new
code by which we can derive photometric evolution of galaxies based on the
detailed spatial distribution of gas and stars and on the time evolution
of star formation rate (Bekki \& Shioya 2000). However, this code cannot
yet enable us to calculate spectroscopic properties of dusty starburst
galaxies due to the time evolution of gaseous emission not yet being
included. Accordingly, an important future task is to derive emission line
properties in numerical simulations. From now on, we refer to these future
numerical simulations as `spectrodynamical simulations', by which we can
investigate not only spectroscopic properties but also morphological, 
structural, and kinematical ones. These will address the following five
questions: (1)\,Can younger starburst populations actually be obscured
more heavily by dust than older ones, and what physical processes are
closely associated with the suggested selective dust extinction?
(2)\,What determines the strength of the age-dependence of $A_V$
in dusty starburst galaxies and is the strength time-dependent? 
(3)\,How do physical processes specific to cluster galaxy evolution,
such as galaxy harassment (Moore et al. 1994) and ram-pressure stripping
(Farouki \& Shapiro 1984), determine the relative spatial distribution of
dusty gas and young stars? 
(4)\,Are supernovae feedback effects important for explaining the
spectroscopic properties of less luminous galaxies with e(b) spectra? 
(5)\,What is the most important dynamical process for the origin of each
of the spectral classes observed in clusters? 

\section{Conclusions}

We have investigated the evolution in spectral properties of dusty
starburst galaxies, adopting the fundamental assumption that the 
amount of dust extinction (quantified by $A_V$) depends on the age of
the stellar population according to:  
$A_V=A_{V,0} \exp \{ (T-t)/\tau \} \; \; \;$, 
where $A_{V,0}$ and $\tau$ are parameters controlling the 
degree of extinction and $T$ and $t$ represent the age of the galaxy
and that of its stellar population, respectively. 
The main conclusions from this investigation can be summarized as follows: 

\begin{enumerate}
\item If the young stellar populations (with an age of $\sim 10^6$\,yr)
are more heavily obscured by dust than the older ones (age$>10^8$\,yr)
in a dusty starburst galaxy, the integrated spectrum of the galaxy 
is an e(a) type characterized by strong H$\delta$ absorption and
relatively modest [OII] emission. The time interval over which the
galaxy exhibits an e(a) spectrum is fairly long, lasting for 
$\sim 0.7$\,Gyr of the $\sim 1$\,Gyr duration of the starburst.

\item There are requisite physical conditions for a dusty starburst
galaxy to show an e(a) spectrum. For example, the mean $A_{V,0}$ should be
within a certain range ($\sim 1<A_{V,0}<\sim 10$). 
The age dependence of dust extinction  
also should be fairly strong in a sense
that young stars with the ages of $\sim$ $10^6$ yr are  preferentially
obscured by dust. 
(i.e., $\tau$ $\sim$ $10^6$ yr; more selective).

\item A dusty starburst galaxy with an e(a) spectrum will evolve into a
poststarburst galaxy with an a+k (or k+a) spectrum, 0.2\,Gyr after the
starburst has ended, and then into one with a passive k-type spectrum, 
1\,Gyr after the starburst. This result clearly demonstrates an
evolutionary link between the following spectral types: e(a), a+k, k+a,
and k.

\item The evolutionary path of a dusty starburst galaxy on the 
EW([OII])--EW(H$\delta$) plane depends on the history of star formation 
within the disk before the starburst, the epoch, duration, and strength of
the starburst, plus the adopted age-dependence of $A_V$. For example,
galaxies of later Hubble type (i.e. with smaller  $k_{\rm disk}$)
are more likely to pass through the bottom-right part of the e(a) region
during a starburst.

\item For a plausible star formation history, a starburst galaxy with the
earlier burst epoch  is more likely to show slightly lower EW([OII]) and
slightly higher EW(H$\delta$) strengths. This result implies that a
dusty starburst at higher redshift is likely to be located in the
bottom-right part of the e(a) region on the EW([OII])--EW(H$\delta$) plane. 

\item A dusty starburst galaxy can show an a+k or k+a spectrum even
in the starburst phase, if the age dependence of $A_V$ is weak 
($\tau\sim 10^7$) for a fixed $A_{V,0}$. This result implies that if
young massive stars with different ages ($<10^7$) are uniformly
obscured by dust, the galaxy shows an a+k or k+a spectrum.

\item The duration of a starburst can control the spectral type evolution
such that a dusty galaxy with a longer starburst time-scale shows a k+a
spectrum rather than an e(a) spectrum during the starburst phase. 
This result, combined with the previous point implies that the origin of
possible starburst  k+a galaxies detected by VLA radio observations (Smail
et al. 1999) can result either from the weak age-dependent dust extinction
or from the relatively long starburst time scales in these dusty
starburst galaxies.

 
\end{enumerate}

These results clearly demonstrate that the  dependence
of  $A_V$  on  ages of young starburst populations
is a crucially important factor for determining
spectral types of galaxies.
We furthermore stress  that the age dependence can be 
closely associated
with the relative distribution of dusty gas and young stars for  dusty
starburst galaxies in distant clusters. 
We will investigate the importance of the relative spatial distribution
of gas and stars in determining spectroscopic properties of galaxies
by using spectrodynamical simulations in our future works.

\acknowledgments

We are  grateful to the referee Michael Dopita for valuable comments,
which contribute to improve the present paper.
Y.S. thanks the Japan Society for Promotion of Science (JSPS) 
Research Fellowships for Young Scientist.
K.B. acknowledges the Large Australian Research Council (ARC).

\newpage
\appendix

\section{A metallicity dependence of [OII]/H$\beta$}

We used the [OII]/H$\beta$ emission line ratio (adopting a value of 3.01
throughout this paper; listed in the table of the PEGASE paper published
by Fioc \& Rocca-Volmerange 1997) to calculate the [OII] emission line
strength of a dusty starburst galaxy from the derived H$\beta$ line
strength in this work. It is important for us to confirm whether or
not our results, described in the main body of the paper, change
if we adopt a [OII]/H$\beta$ ratio different to that of the PEGASE
model. We therefore demonstrate here that our results do not depend
strongly on the adopted [OII]/H$\beta$ value, by showing results 
calculated by the photoionization code CLOUDY (Ferland et al. 1998).   

Before presenting the results, we confirm whether our very simple
model with the CLOUDY code can reproduce reasonably well the
[OII]/H$\beta$ values observed for nearby galaxies. Figure 11 shows
dereddened [OII]/H$\beta$ ratios as a function of oxygen abundance  
($12 + \log ({\rm O/H}$) from the observations of Kobulnicky et
al. (1999). In this figure, we also give our theoretical results based on
the CLOUDY code, with an SED of continuous star formation of 10 Myr,
an electron density of $10^3$\,cm$^{-3}$, and an ionizing parameter $U$
of $10^{-3}$, for different metallicity models. As is shown in this
figure, the CLOUDY--based model can successfully reproduce the
observational results for galaxies with $12 + \log ({\rm O/H}$ $<$ 8.5.
It should be noted here that the [OII]/H$\beta$ ratio is less than 3
(corresponding to the value in the PEGASE with solar metallicity) for  
$8.5 < 12 + \log {\rm O/H}$ and has a value of $\sim 2$ around solar
metallicity ($12 + \log {\rm O/H} \sim 8.9$). This result suggests that
the [OII] emission line strength calculated by PEGASE is appreciably
larger (by a factor of $\sim 1.5$) than that of CLOUDY.

Considering the above difference in the [OII]/H$\beta$ ratio between the
two models, we investigate the evolution of a dusty starburst galaxy on
the EW([OII])-EW(H$\delta$) plane for Models 1 and 2 by using the CLOUDY
code with solar-metallicity (the corresponding [OII]/H$\beta$ $\sim$
value is 2.08). As is shown in Figure 12, even if we use the CLOUDY code,
our key results on e(a) formation and spectral evolution do not change
(compare Figure 4 with Figure 12). However, the two models based on the
CLOUDY code show lower EW([OII]) than the PEGASE-based models
during their evolution. This is a result of the smaller [OII]/H$\beta$
ratio (2.08), which in turn makes EW([OII]) smaller.
As a natural results of this, the model with dust shows an e(c) rather
than an e(b) spectrum during the actively star-forming phase just prior
to the starburst. 

For all of the present models, we assumed that the [OII]/H$\beta$ ratio is 
fixed at a certain value for all time steps in a model. This is actually
an over-simplification, because the gaseous metallicity (thus
[OII]/H$\beta$ ratio) in a galaxy evolves with time as the chemical
evolution of the galaxy proceeds. Owing to this simplification, the
present models show larger values of EW([OII]) in the early evolution
phase. In future work, we will consider the dependence of the
[OII]/H$\beta$ ratio on the gaseous metallicity and thereby construct a
more realistic and sophisticated model of the spectral evolution of dusty
galaxies.

Lastly, we point out one disadvantage of the present model in deriving
the [OII] line of dusty starburst galaxies. Since we used the PEGASE code
for the calculation of [OII]/H$\beta$, 
the temperature of the HII region in the present model with the PEGASE code 
is  too low, in particular, for high metallicity, owing to the model's
not allowing for any dust depletion onto grains in the HII regions.
As a natural result of this, the [OII] line strength in the present model
is severely decreased relative to what it should be.
One of the best way to avoid this underestimation of [OII] line strength
is to use the [OII]/H$\beta$ derived by Dopita et al. (2000) in which
dust physics and the depletion of various elements out of the gas phase 
are more properly treated. 
We have a plan to investigate the [OII]/H$\beta$ of dusty starburst galaxies
in our future papers by using
the model by Dopita et al. (2000).


\clearpage

\figcaption{The star formation history ({\it upper panel})
and the $B-V$ color evolution ({\it lower panel}) for the fiducial model
(Model 1) with the Sb-type star formation. For comparison, the color
evolution for the model {\it without} dust extinction is also
given in the {\it lower panel}. The starburst begins at $T_{sb}=7.64$\,Gyr
and stops at $T_{end}=8.56$\,Gyr. We assume here that the mass of the
model galaxy is $10^{11}M_{\odot}$ (corresponding to the disk mass of the
Galaxy).
\label{fig-1}}

\figcaption{The spectral energy distribution (SED) with emission lines
at $T = 1.0$ ({\it top}), 7.64 ({\it second from the top}), 8.56 
({\it middle}), 8.58 ({\it second from bottom}), and 13.0\,Gyr
({\it bottom}) for the fiducial model. The continuum and the emission
lines are represented by the {\it solid} and {\it dotted} lines,
respectively. The time $T = 7.64$, 8.56, and 8.58 represent the beginning
of the starburst, the end of the starburst, and the epoch when
EW([OII]) reaches a minimum, respectively.
\label{fig-2}}

\figcaption{{\it Upper panel}: the evolution of the equivalent width of
the [OII] emission line, [EW([OII])], for the fiducial model. 
{\it Lower panel}: The evolution of the equivalent width of the H$\delta$
absorption line, [EW(H$\delta$)]. The {\it solid}, {\it dotted}, and
{\it dashed} lines represent the equivalent width of the stellar absorption, 
the equivalent width of emission lines formed in ionized gas, and the sum
of both lines, respectively. Positive (negative) EW(H$\delta$) means that
the H$\delta$ line is observed in absorption (emission). 
\label{fig-3}}

\figcaption{The evolution of galaxies on the EW([OII])--EW(H$\delta$) plane 
for the fiducial model ({\it upper panel}) and the model without dust
extinction (Model 2). Some points ($T=1, 7.64, 8.56, 8.58, 12$\,Gyr) are
shown as {\it open circles} (Model 1) and {\it open triangles} (Model 2) 
along the loci of the two models. The classification criteria of
Dressler et al. (1999) are also superimposed. Both of the models evolve
from the e(b) region ($T = 1.0$\,Gyr) into the k region ($T = 12$\,Gyr).
Note that only the fiducial model can pass through the e(a) region 
during the starburst ($7.64<T<8.56$\,Gyr), which implies that the
selective dust extinction during the starburst is very important for the
formation of e(a) galaxies. 
\label{fig-4}}

\figcaption{The same as Figure 4 but for the models with different 
$A_{V,0}$. The results for Model 3 with $A_{V,0}$ = 1.0,
Model 4 with $A_{V,0}$ = 3.0, Model 1 with $A_{V,0}$ = 5.0,
and Model 5 with $A_{V,0}$ = 10.0 are represented by the {\it dotted} 
line with {\it open triangles}, {\it short-dashed} line with {\it squares}, 
{\it solid} line with {\it circles}, and {\it long-dashed} line with
{\it pentagons}, respectively. The results for the five important time
steps at $T$ = 1.0, 7.64, 8.56, 8.58, and 13.0 Gyr are indicated by
the {\it open circles}.
\label{fig-5}}

\figcaption{As for Figure 5 but for the models with different $\tau$.
The results for Model 1 with $\tau = 1.0 \times 10^6$, Model 6 with
$\tau = 3.0 \times 10^6$, and Model 7 with $\tau = 1.0 \times 10^7$
are plotted as a {\it solid} line with {\it circles}, a {\it dotted}
line with {\it open triangles}, and a {\it short-dashed} line with
{\it squares}, respectively.
\label{fig-6}}

\figcaption{{\it Top}: The time evolution of the star formation rate for
models with different $T_{\rm sb}$. {\it Bottom}: The same as Figure 5
but for the models with different $T_{\rm sb}$. The results for Model 8
with $T_{\rm sb} = 4.46$, Model 9 with $T_{\rm sb} = 6.93$, Model 1 with
$T_{\rm sb} = 7.64$, Model 10 with $T_{\rm sb} = 8.47$, and Model 11 with
$T_{\rm sb} = 9.45$ are represented by a {\it dotted} line with 
{\it open triangles}, a {\it short-dashed} line with {\it squares}, a 
{\it solid} line with {\it circles}, a {\it long-dashed} line with
{\it pentagons}, and a {\it dot-dashed} line with {\it crosses}, 
respectively.
\label{fig-7}}

\figcaption{As for Figure 7 but for different $k_{\rm disk}$ values. 
The results for Model 12 with $k_{\rm disk} = 0.056$ (Sc type),
Model 1 with $k_{\rm disk} = 0.225$ (Sb type), and Model 13 with 
$k_{\rm disk} = 0.325$ (Sa type), are represented by a {\it dotted} line
with {\it open triangles}, a {\it solid} line with {\it circles}, and
a {\it short-dashed} line with {\it squares}, respectively.
\label{fig-8}}

\figcaption{As for Figure 7 but for different $k_{\rm sb}$ values.
The results for Model 14 with $k_{\rm sb} = 1.0$, Model 1 with 
$k_{\rm sb} = 0.225$, and Model 15 with $k_{\rm sb} = 5.0$,  
are represented by a {\it dotted} line with {\it open triangles}, a
{\it solid} line with {\it circles}, and a {\it short-dashed} line with
{\it squares}, respectively.
\label{fig-9}}

\figcaption{
The same as Figure 7 but for different $f_{\rm final}$ values. The results
for Model 16 with $f_{\rm final} = 0.90$, Model 1 with $k_{\rm final} =
0.95$, and Model 17 with $f_{\rm final} = 0.99$, are represented by 
a {\it dotted} line with {\it open triangles}, a {\it solid} line with
{\it circles}, and a {\it short-dashed} line with {\it squares}, 
respectively.
\label{fig-10}}

\figcaption{The observed relation between the [OII]/H$\beta$ ratio and
oxygen abundance ($12 + \log {\rm O/H}$) for spiral galaxies ({\it open
circles}) and metal-poor irregular dwarf galaxies ({\it crosses}) from  
Kobulnicky et al. (1999). The {\it solid} line represents our numerical
results based on  CLOUDY94 (See the text for more details). 
\label{fig-11}}

\figcaption{
The same as Figure 4 but for the different [OII]/H$\beta$ ratio derived from CLOUDY94 
(See the text for more details). 
\label{fig-12}}


\clearpage

\begin{deluxetable}{cccccccc}
\footnotesize
\tablecaption{Model parameters \label{tbl-1}}
\tablewidth{0pt}
\tablehead{
\colhead{model} & \colhead{$k_{\rm disk}$} & \colhead{$T_{sb}$} 
& \colhead{$k_{\rm sb}$} & \colhead{$f_{\rm final}$} 
& \colhead{$T_{\rm end}$} & \colhead{$\tau$} & \colhead{$A_{V,0}$} \\
\colhead{} & \colhead{(${\rm Gyr}^{-1}$)} & \colhead{(Gyr)} 
& \colhead{(${\rm Gyr}^{-1}$)} & \colhead{} 
& \colhead{(Gyr)} & \colhead{(yr)} & \colhead{(mag)}}
\startdata
1  & 0.225& 7.64& 2.25& 0.95& 8.558& $1 \times 10^6$ & 5 \\
2  & 0.225& 7.64& 2.25& 0.95& 8.558& $1 \times 10^6$ & 0\\
3  & 0.225& 7.64& 2.25& 0.95& 8.558& $1 \times 10^6$ & 1 \\
4  & 0.225& 7.64& 2.25& 0.95& 8.558& $1 \times 10^6$ & 3 \\
5  & 0.225& 7.64& 2.25& 0.95& 8.558& $1 \times 10^6$ & 10\\
6  & 0.225& 7.64& 2.25& 0.95& 8.558& $3 \times 10^6$ & 5 \\
7  & 0.225& 7.64& 2.25& 0.95& 8.558& $1 \times 10^7$ & 5 \\
8  & 0.225& 4.46& 2.25& 0.95& 5.722& $1 \times 10^6$ & 5 \\
9  & 0.225& 6.93& 2.25& 0.95& 7.920& $1 \times 10^6$ & 5 \\
10 & 0.225& 8.47& 2.25& 0.95& 9.307& $1 \times 10^6$ & 5 \\
11 & 0.225& 9.45& 2.25& 0.95&10.193& $1 \times 10^6$ & 5 \\
12 & 0.056& 7.64& 2.25& 0.95& 9.189& $1 \times 10^6$ & 5 \\
13 & 0.325& 7.64& 2.25& 0.95& 8.242& $1 \times 10^6$ & 5 \\
14 & 0.225& 7.64& 1.00& 0.95& 9.870& $1 \times 10^6$ & 5 \\
15 & 0.225& 7.64& 5.00& 0.95& 8.030& $1 \times 10^6$ & 5 \\
16 & 0.225& 7.64& 2.25& 0.90& 8.148& $1 \times 10^6$ & 5 \\
17 & 0.225& 7.64& 2.25& 0.99& 9.921& $1 \times 10^6$ & 5 \\
\enddata
\end{deluxetable}

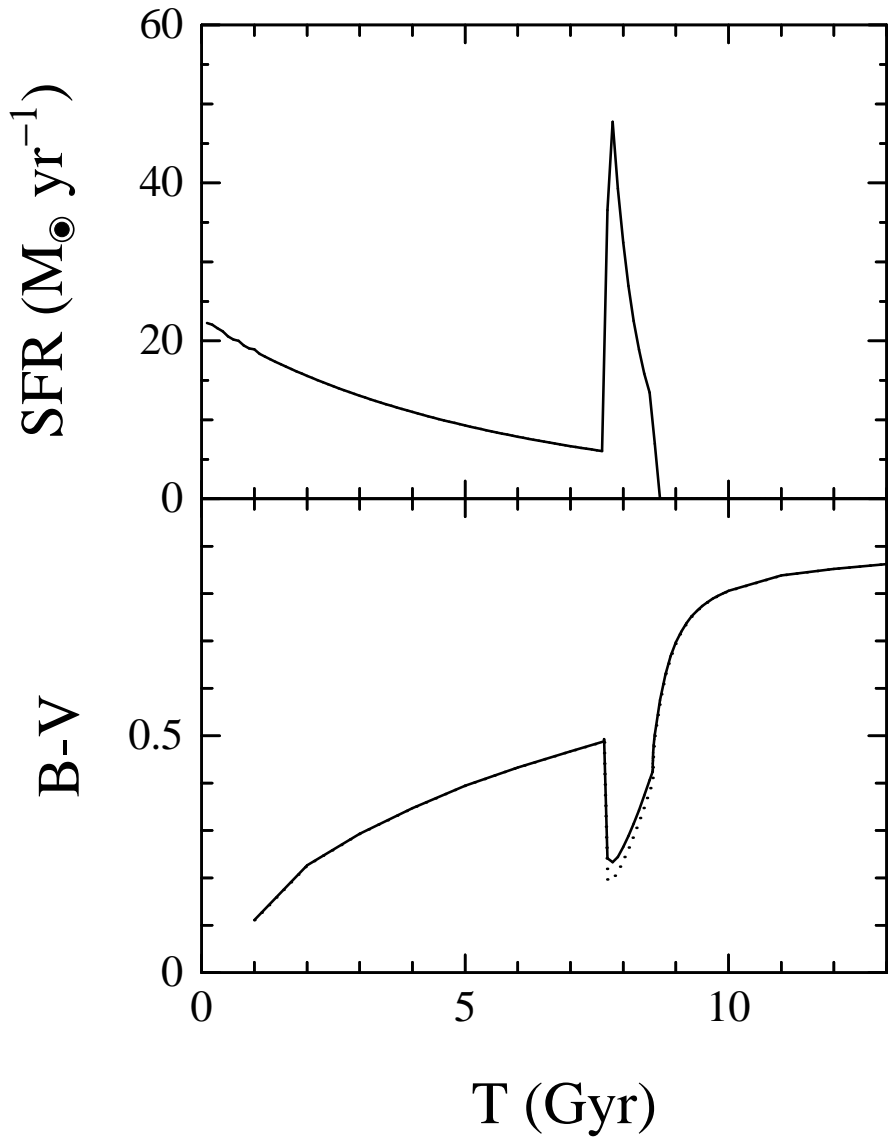
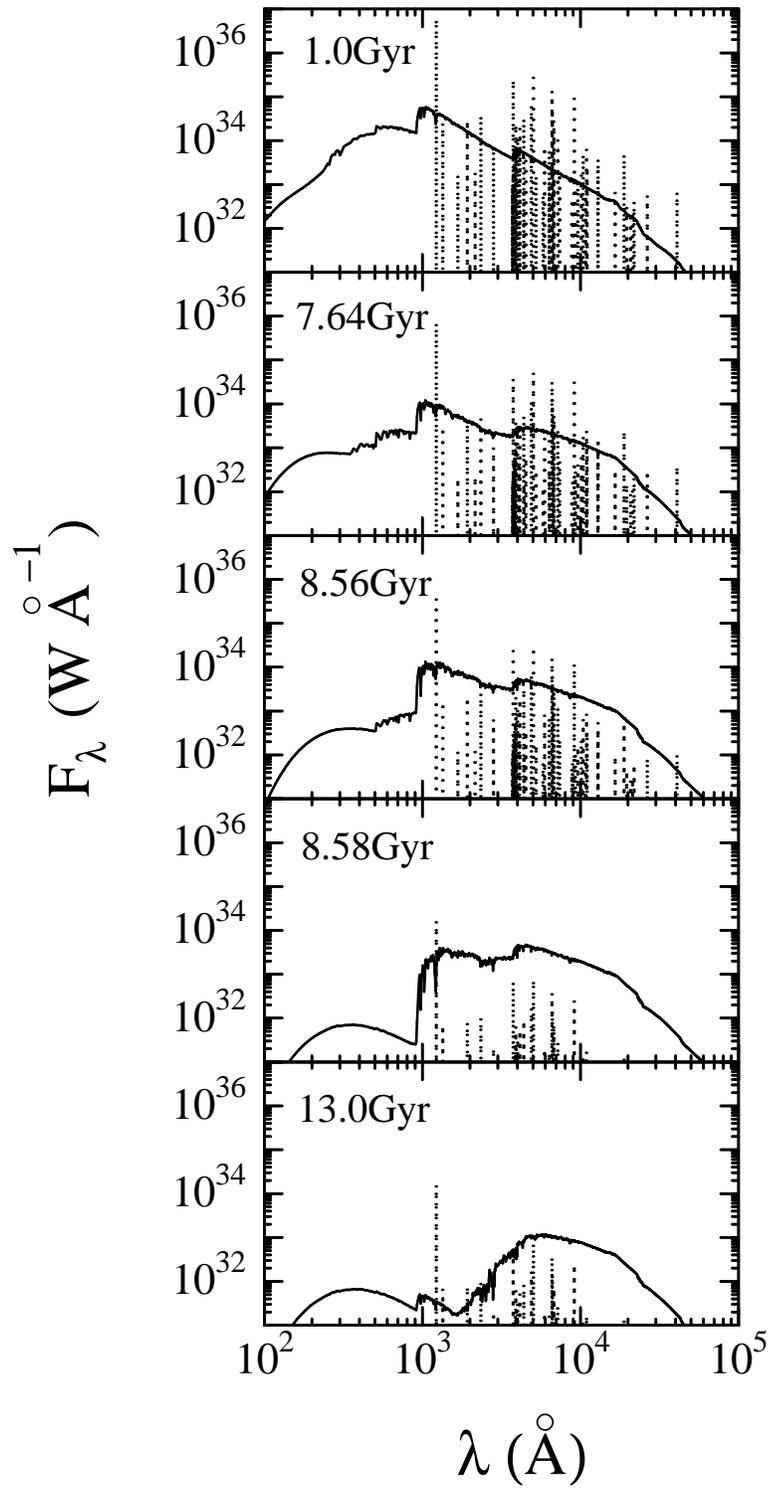
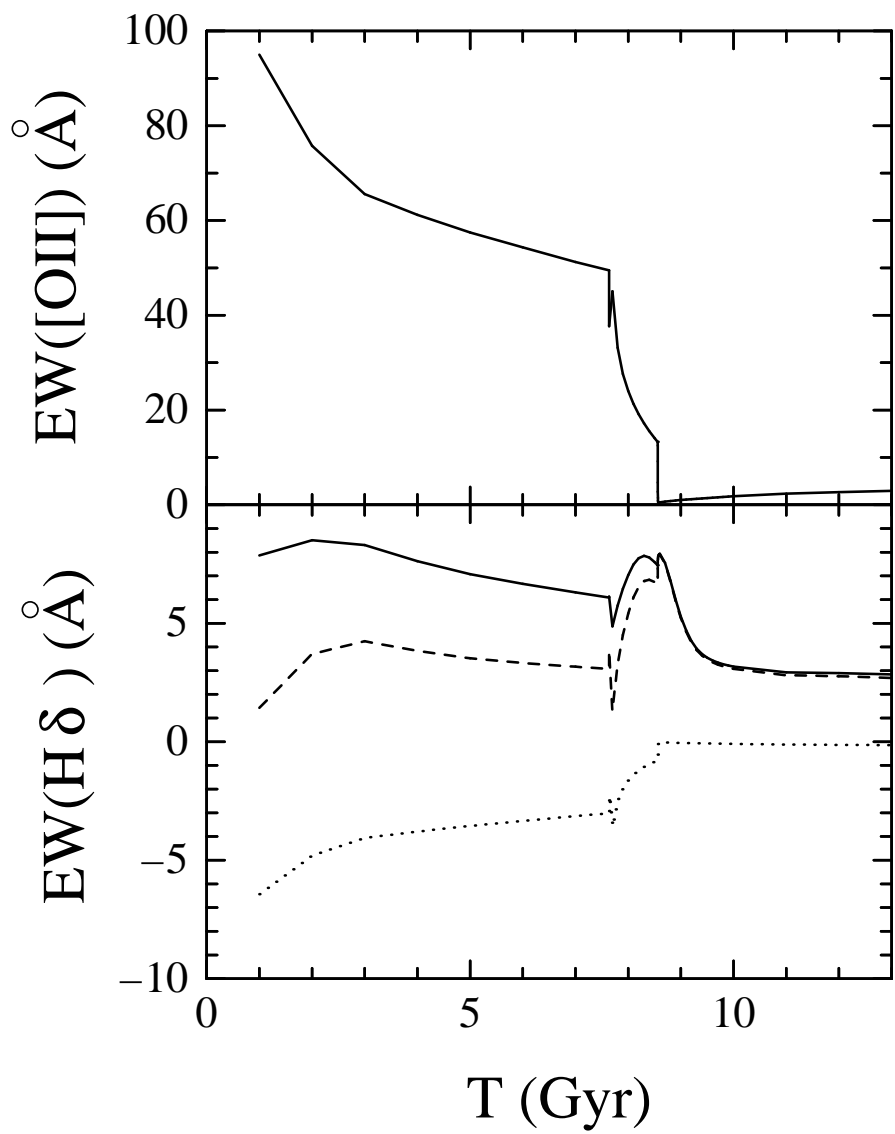
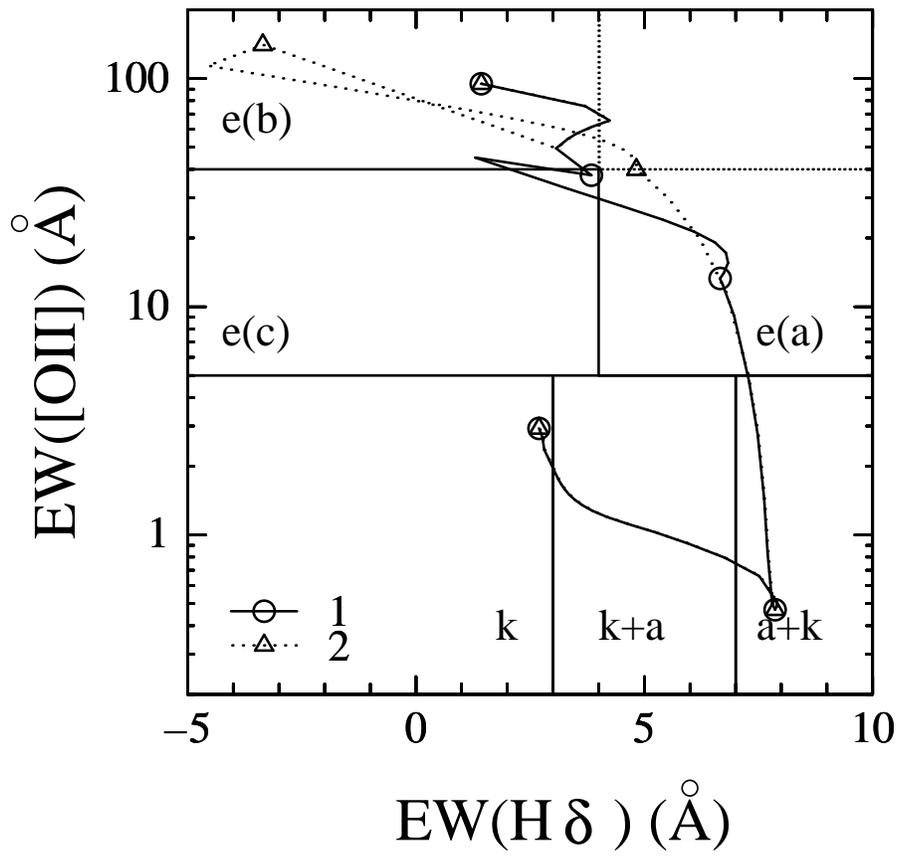
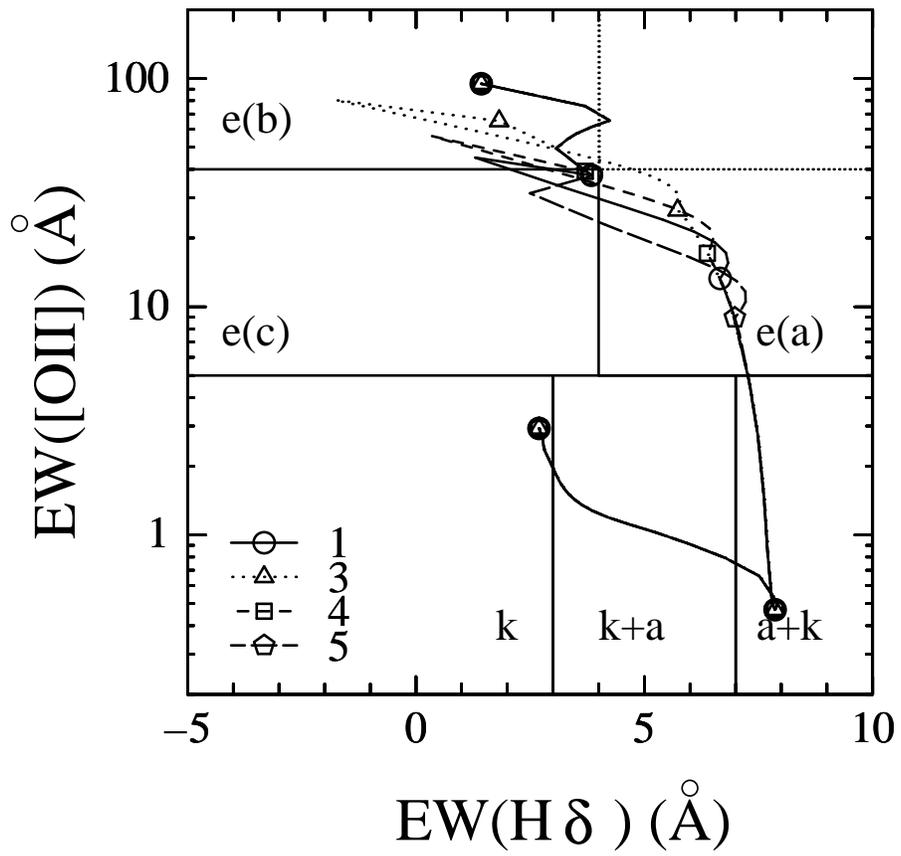
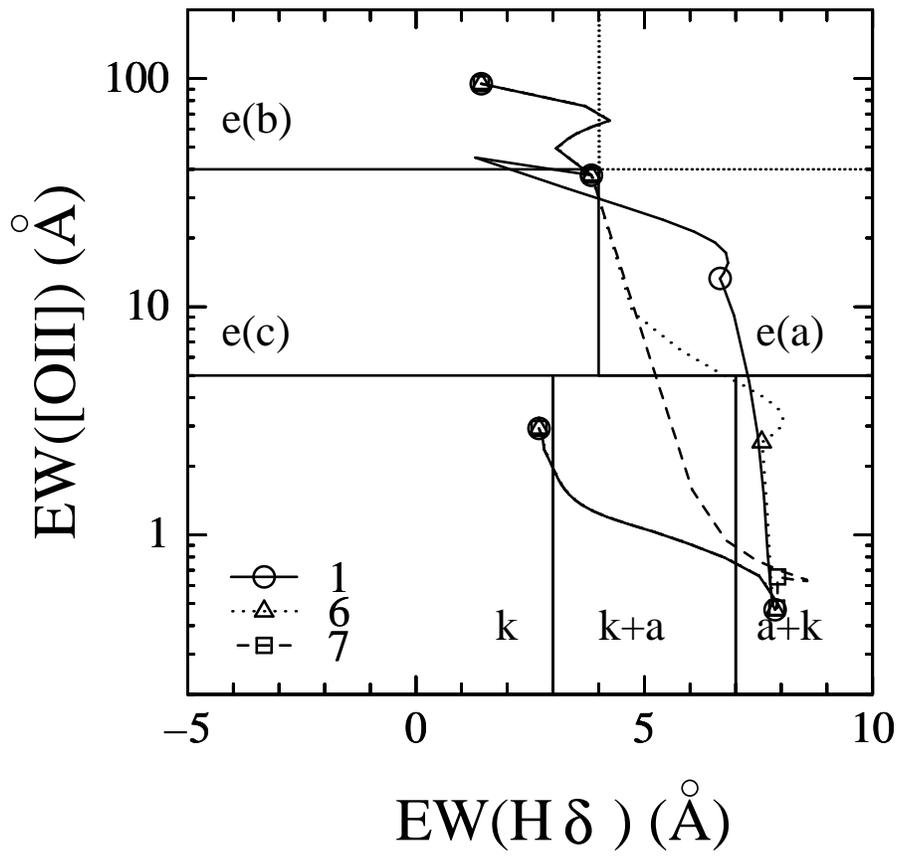
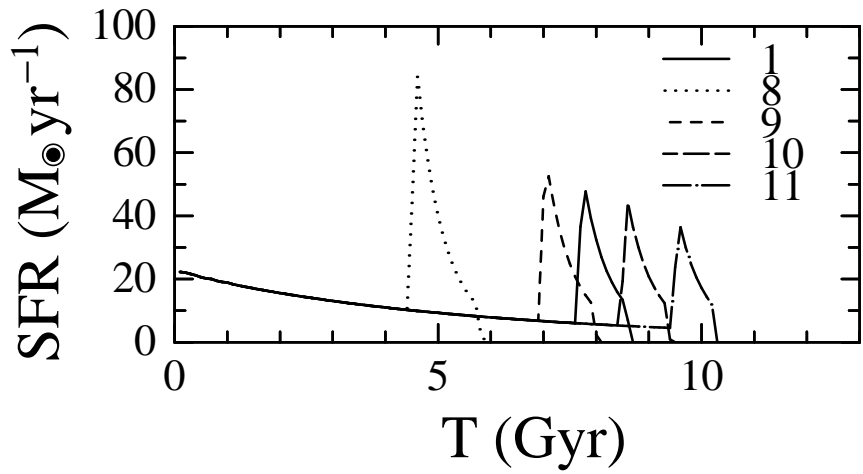
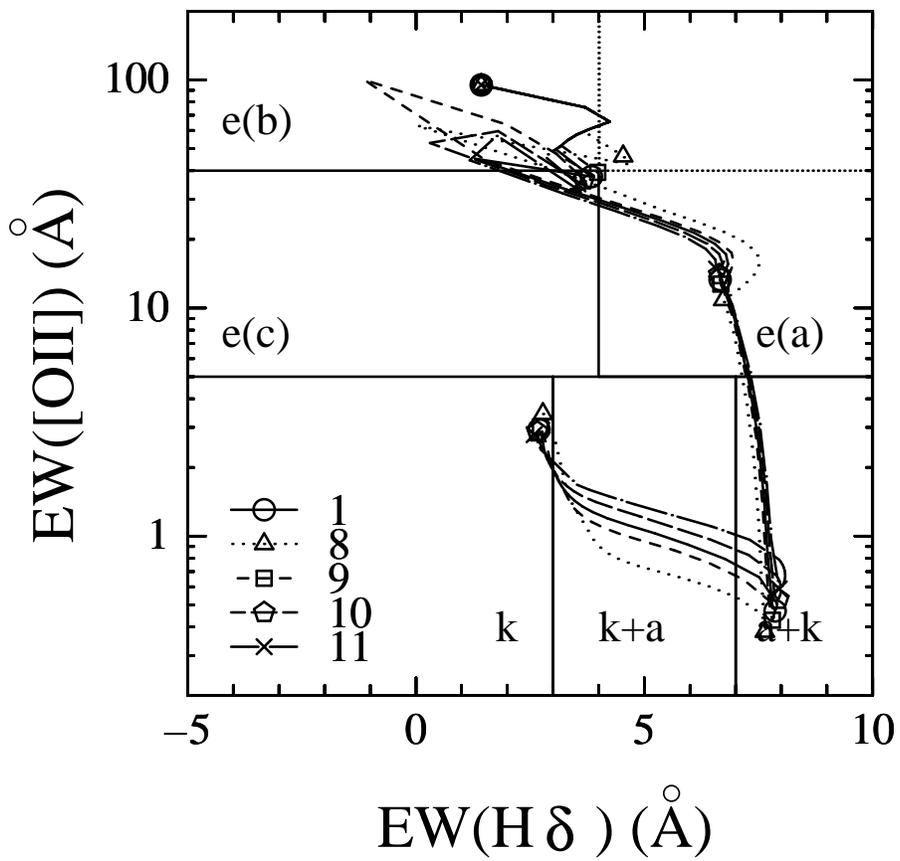
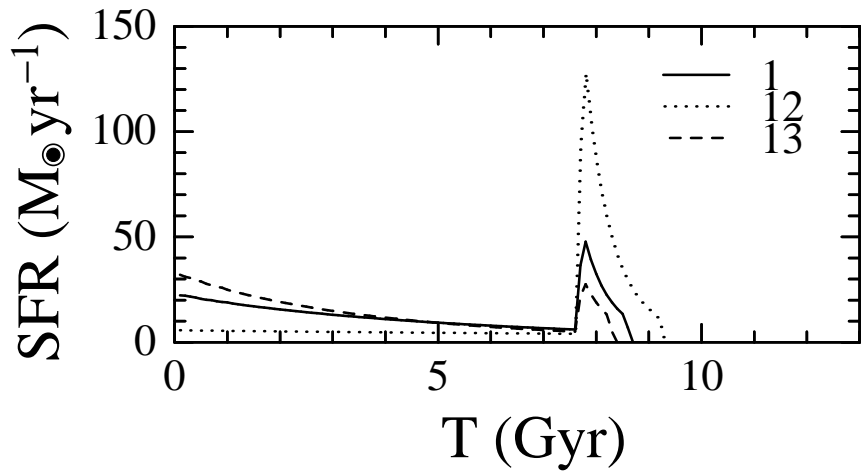
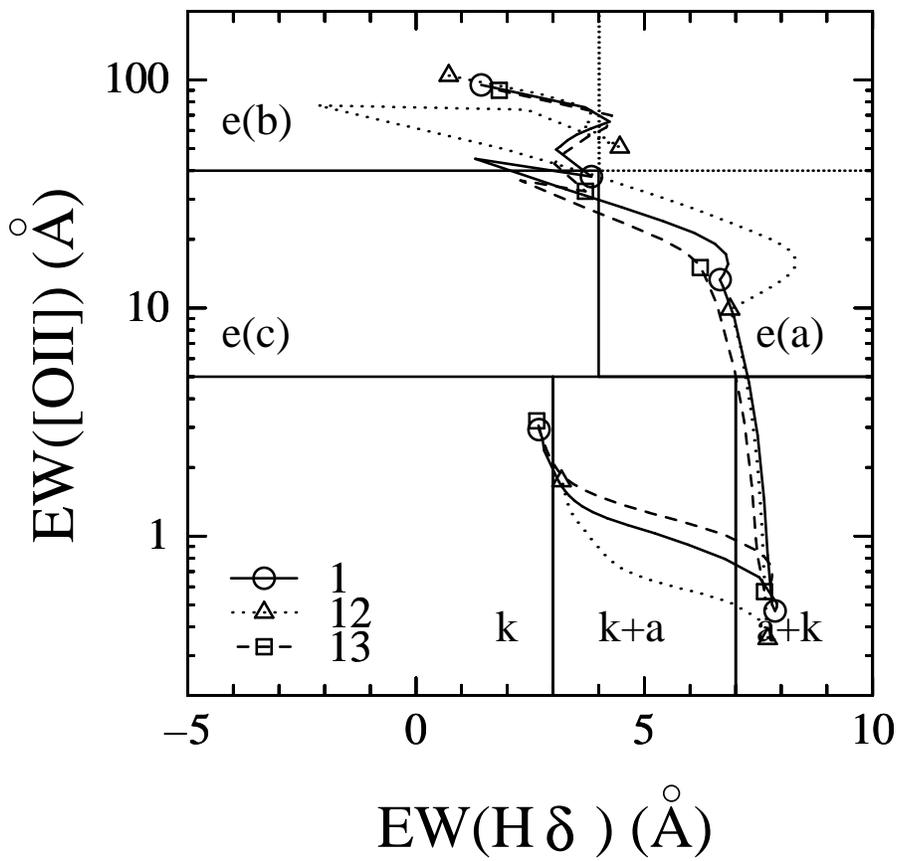
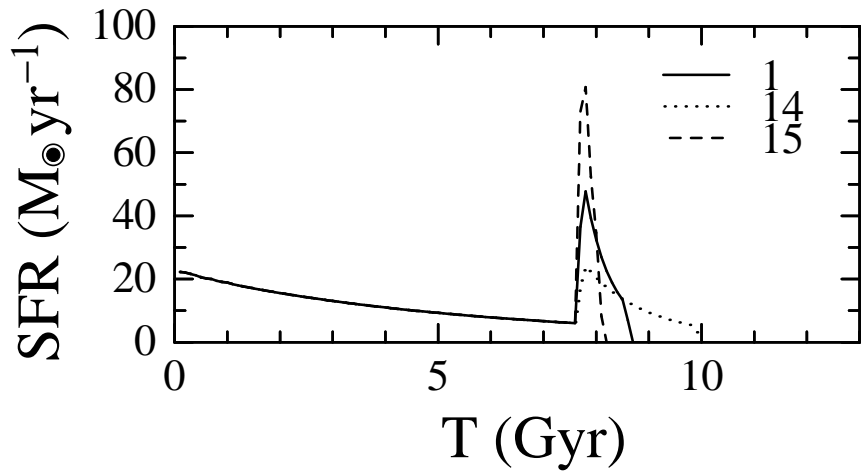
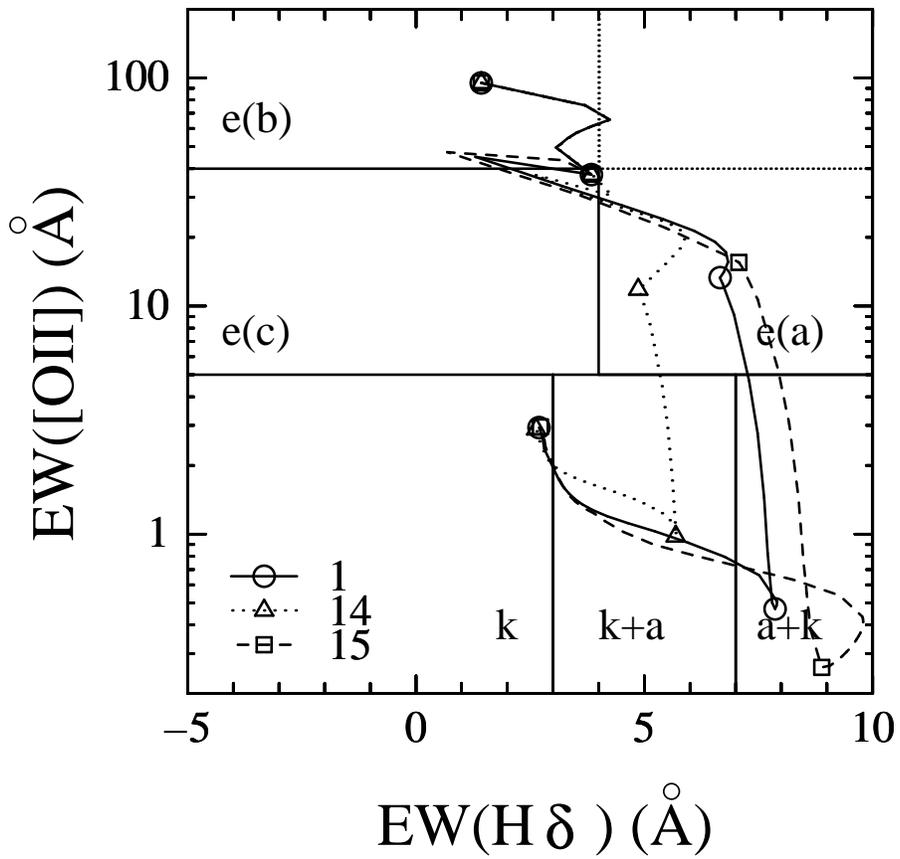
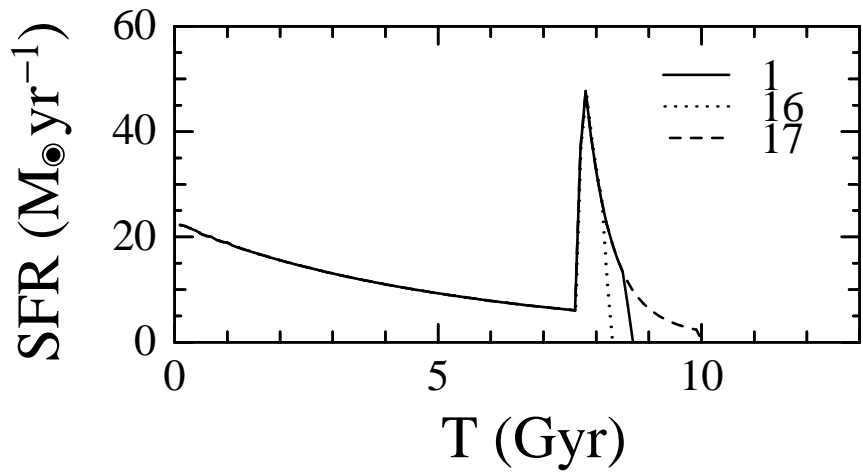
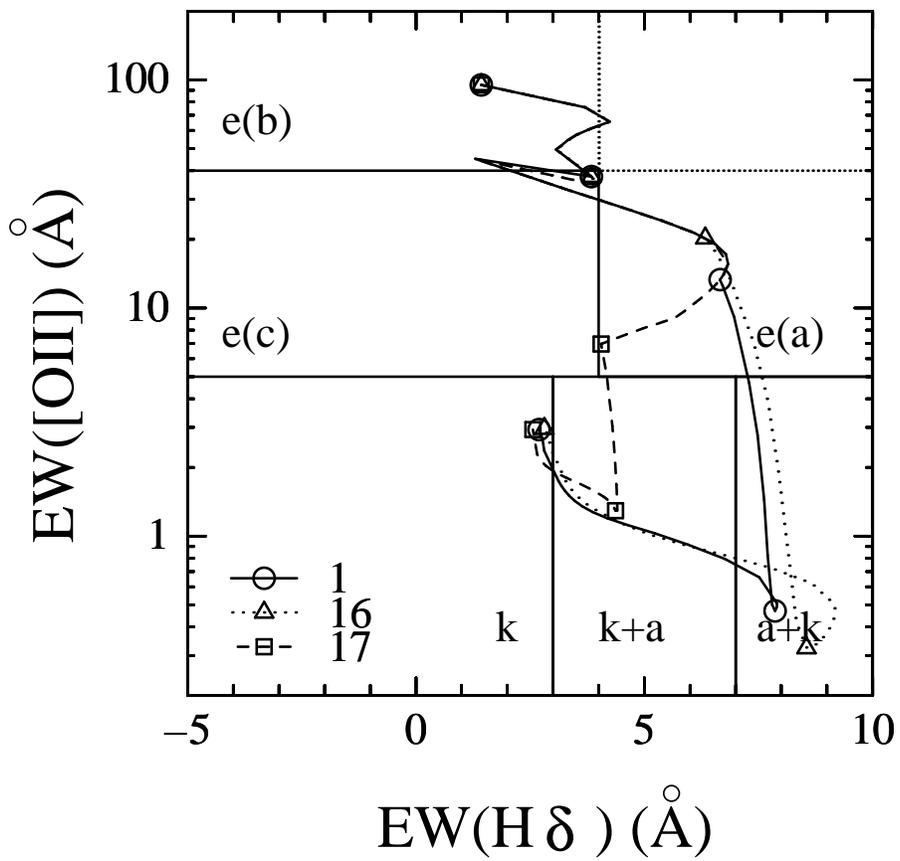
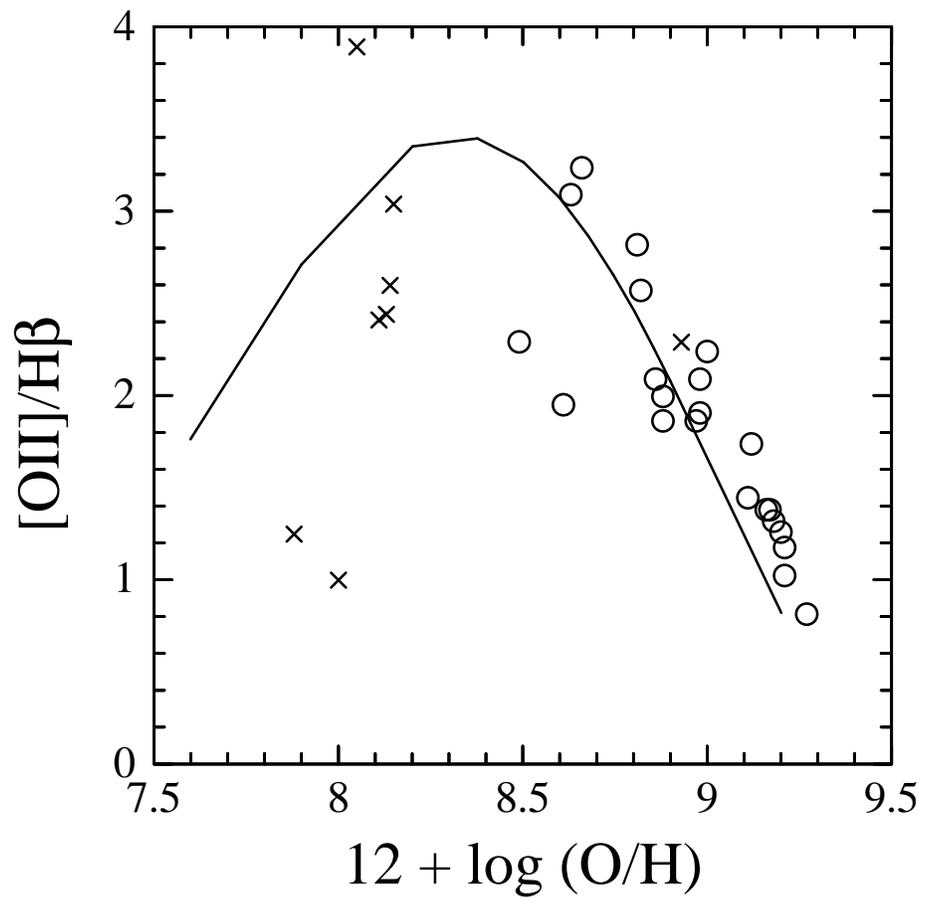
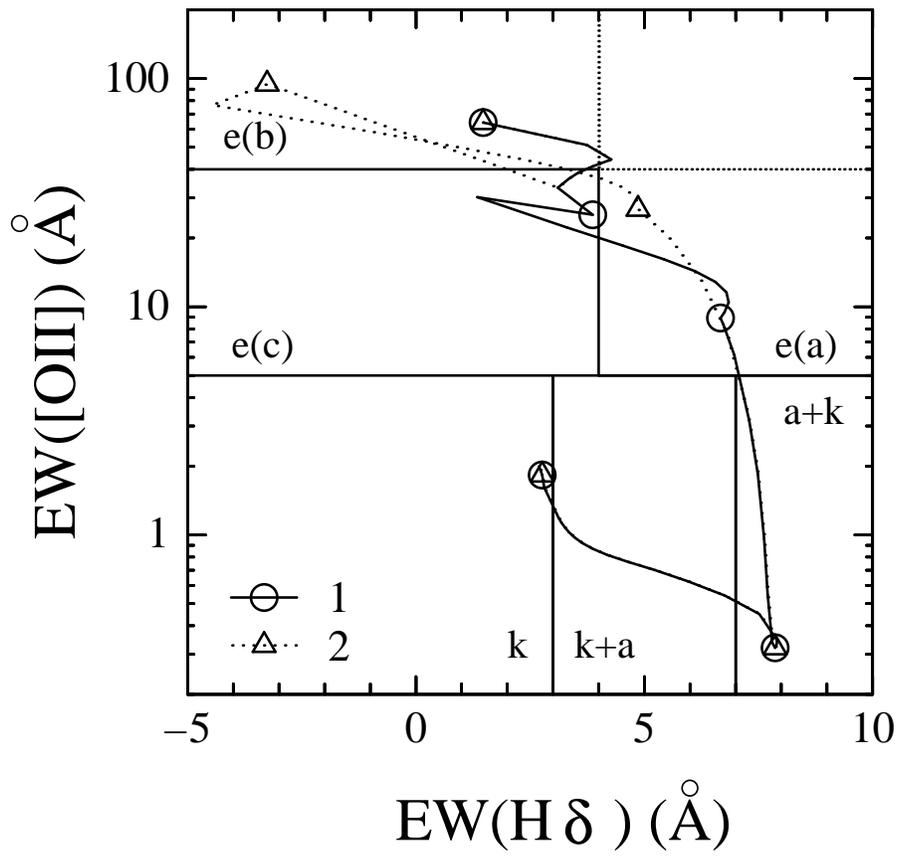
\end{document}